\documentclass[%
reprint,
superscriptaddress,
amsmath,amssymb,
aps,
prl,
floatfix,
]{revtex4-1}
\usepackage{bm}
\usepackage{physics}
\usepackage{graphicx}
\usepackage{float}
\usepackage{booktabs}
\usepackage{multirow}
\usepackage{xcolor}

\begin{document}

\title{Chemically-Localized Resonant Excitons in Silver-Pnictogen Halide Double Perovskites}

\author{Raisa-Ioana Biega}
\affiliation{Institute of Physics, University of Bayreuth, Bayreuth 95440, Germany}

\author{Marina R. Filip}
\email{marina.filip@physics.ox.ac.uk}
\affiliation{Department of Physics,  University of Oxford, Clarendon Laboratory, Oxford,  OX1 3PU, United Kingdom}
\affiliation{Department of Physics, University of California, Berkeley, CA 94720, USA}
\affiliation{Materials Science Division,Lawrence Berkeley National Laboratory, Berkeley, CA 94720, USA}

\author{Linn Leppert}
\email{l.leppert@utwente.nl}
\affiliation{MESA+ Institute for Nanotechnology, University of Twente, 7500 AE Enschede, The Netherlands}
\affiliation{Institute of Physics, University of Bayreuth, Bayreuth 95440, Germany}

\author{Jeffrey B. Neaton}
\email{jbneaton@lbl.gov}
\affiliation{Department of Physics, University of California, Berkeley, CA 94720, USA}
\affiliation{Materials Science Division,Lawrence Berkeley National Laboratory, Berkeley, CA 94720, USA}
\affiliation{Kavli Energy NanoSciences Institute at Berkeley, Berkeley, CA 94720, USA}
\date{\today}

\bibliographystyle{apsrev4-1}

\begin{abstract}
Halide double perovskites with alternating silver and pnictogen cations are an emerging family of photoabsorber materials with robust stability and band gaps in the visible range. However, the nature of optical excitations in these systems is not yet well understood, limiting their utility.
Here, we use \textit{ab initio} many-body perturbation theory within the $GW$ approximation and the Bethe-Salpeter equation approach to calculate the electronic structure and optical excitations of the double perovskite series Cs$_2$AgBX$_6$, with B=Bi$^{3+}$, Sb$^{3+}$, X = Br$^-$, Cl$^-$.
We find that these materials exhibit strongly localized resonant excitons with energies from 170 to 434 meV below the direct band gap. In contrast to lead-based perovskites, the Cs$_2$AgBX$_6$ excitons are computed to be non-hydrogenic, with anisotropic effective masses and sensitive to local field effects, a consequence of their chemical heterogeneity. Our calculations demonstrate the limitations of the Wannier-Mott and Elliott models for this class of double perovskites and contribute to a detailed atomistic understanding of their light-matter interactions.
\end{abstract}

\maketitle
Halide double perovskites with the stoichiometry A$_2$B'BX$_6$ are an exciting class of optoelectronic materials with great chemical and functional diversity through substitution at the A, B', B and X sites \cite{McClure2016, Volonakis2016, Wei2016,Volonakis2017,Slavney2018-TlAg}. In these compounds, octahedrally coordinated B' and B metal cations occupy alternating lattice sites, allowing for the incorporation of metals with nominal oxidation states from +1 to +4 \cite{Giustino2016ACSrev,Slavney2019}. The compositional flexibility of double perovskites gives rise to a large number of thermodynamically stable materials \cite{Faber2016,Filip2018} with a rich variety of electronic structures; many halide double perovskites have been synthesized and studied as potential solar absorbers \cite{Deng2016c,Debbichi2018,Yang2018angewandte}, X-ray detectors, \cite{Luo2018} and scintillators \cite{Biswas2012}. The double perovskites Cs$_2$AgBX$_6$ (B=Bi, Sb; X=Br, Cl) have received particular attention due to their attractive semiconducting properties. In particular, Cs$_2$AgBiBr$_6$ is highly stable \cite{Wu2018}, has low carrier effective masses \cite{Volonakis2016}, long recombination lifetimes \cite{Slavney2016a}, large carrier mobilities~\cite{Delor2020}, and an indirect band gap of $\sim$2\,eV\cite{McClure2016, Steele2018a}. Despite the relatively large, indirect band gap, Cs$_2$AgBiBr$_6$-based solar cells with $\sim$3\% power conversion efficiency have recently been demonstrated \cite{Greul2017,Yang2020}.

The electronic structure of Cs$_2$AgBiBr$_6$ has been primarily studied using density functional theory (DFT) with standard (semi)local approximations\cite{Volonakis2016, Filip2016}, which are known to substantially underestimate quasiparticle (QP) band gaps. The screened range-separated hybrid functional HSE06 predicts the band gap of Cs$_2$AgBiBr$_6$ to be 1.8\,eV \cite{McClure2016}, but underestimates the band gaps of other double perovskites \cite{Leppert2019l}. More accurate QP band gaps and band structures can be obtained using Green's function-based \textit{ab initio} many-body perturbation theory with the $GW$ approximation. The QP band gap of Cs$_2$AgBiBr$_6$ has been computed with $GW$ approaches to lie between 1.8\,eV~\cite{Filip2016} and 2.2\,eV~\cite{Leppert2019l}, in good agreement with the range of experimental values.

Optical properties of Cs$_2$AgBiBr$_6$ have been investigated both theoretically and experimentally. 
Several experimental studies report prominent spectral features typically associated with excitons. Ref.~\citenum{Bartesaghi2018a} reports a peak with a bandwidth on the order of 200\,meV that is clearly resolved with respect to the onset of the band-to-band absorption and does not significantly change with temperature, indicating that electron-hole pairs are strongly bound. Similar excitonic features are also seen in the optical spectra of related double perovskites, such as Cs$_2$AgSbCl$_6$~\cite{Dahl2019, Tran2017}. Ref.~\citenum{Yang2018angewandte}~reports a well-defined peak in the optical absorption spectrum and substantial trapping of electrons and holes in Cs$_2$AgBiBr$_6$ nanoparticles, consistent with excitonic effects. In addition, similar excitonic features are reported in both thin films and single crystal Cs$_2$AgBiBr$_6$ in optical measurements reported by several groups \cite{Kentsch2018a, Steele2018, Zelewski2019, Longo2020}. The exciton binding energies themselves remain unclear: reports of measured binding energies range between 70\,meV~\cite{Steele2018} and 268\,meV~\cite{Kentsch2018a}, and a calculated exciton binding energy of 340\,meV obtained with the $GW$+Bethe-Salpeter equation (BSE) approach falls outside this range~\cite{Palummo2020}. This calls for a systematic study of the nature of excitons and their binding energies in this class of materials.

In this Letter, we compute the band structure and excitonic properties of the Cs$_2$AgBX$_6$ series (with B = Bi and Sb and X = Cl and Br) using \textit{ab initio} many-body perturbation theory. We compare our calculations with recent measurements of the optical absorption spectrum of Cs$_2$AgBiBr$_6$ thin films, and assign the well-defined peak at the onset of the optical spectrum to a resonant excitonic feature. We find that Bi- and Sb-based halide double perovskites exhibit localized non-hydrogenic resonant excitons, with energies between 170 and 434~meV below the direct band gap. We show that the departure of these excitons from the hydrogenic Wannier-Mott picture can be explained via an anisotropic effective mass and significant local field effects, a consequence of the local chemical heterogeneity of these materials. The degree of exciton localization correlates with the fractional pnictogen character of the conduction band edge states.

\begin{figure}[htb]
  	\centering
  	\includegraphics[width=0.9\columnwidth]{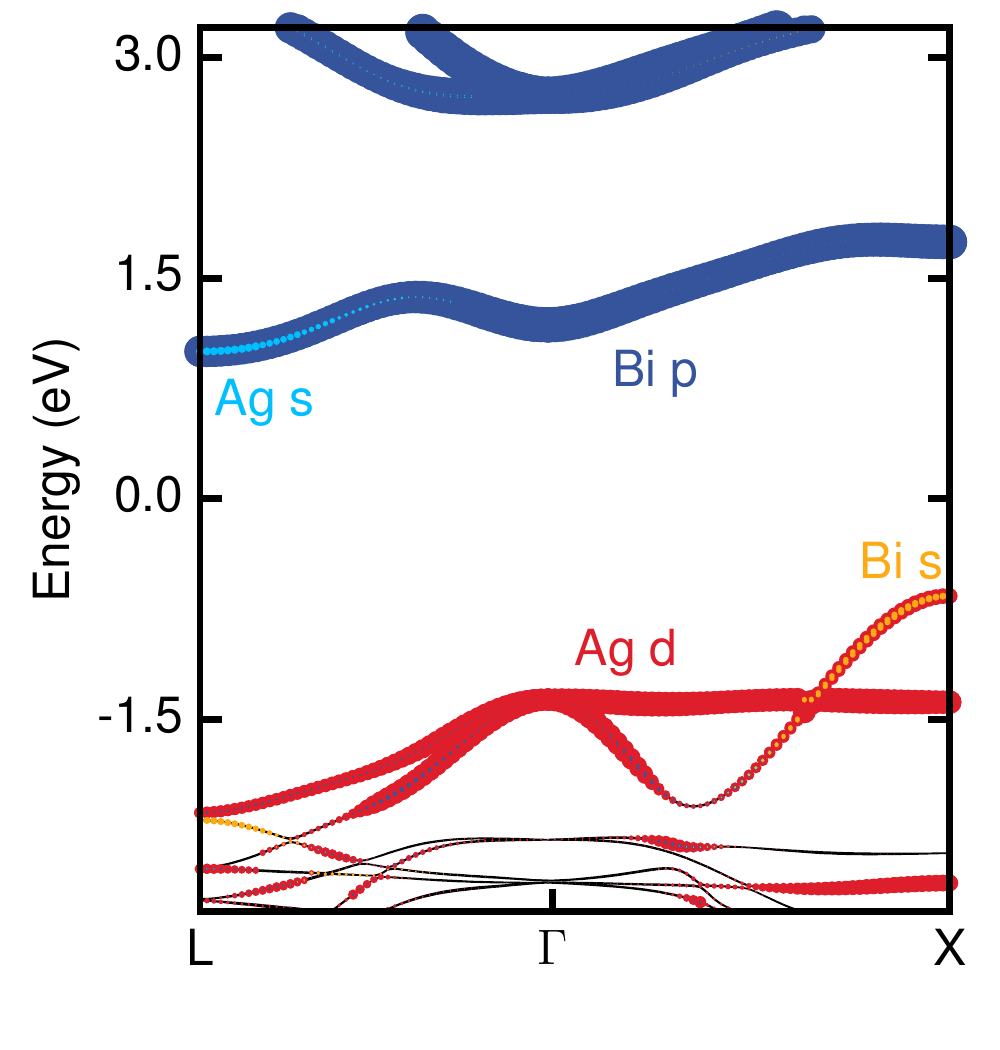}
  	\caption{Quasiparticle band structure of cubic $Fm\bar{3}m$ Cs$_2$AgBiBr$_6$ along the L $[1/2,1/2,1/2]2\pi/a$ - $\Gamma$ $[0,0,0]$ - X $[0,1,0]2\pi/a$ path. Colors represent the orbital character of the bands. The size of the colored dots is proportional to the percentage contribution of the orbital character to the electronic bands. Cs-derived orbitals do not contribute to the states near the band edges and halide character was omitted for clarity.}
  	\label{fig1}
  \end{figure}

In Figure~\ref{fig1}, we show the QP band structure of cubic Cs$_2$AgBiBr$_6$, calculated within the $GW$ approximation~\cite{Hybertsen1987}, overlayed with the orbital character of the energy bands. We use a one-shot $G_0W_0$ approach, in which the QP eigenvalues $E_{nk}^{\text{QP}}$ are calculated by perturbatively correcting DFT-LDA Kohn-Sham eigenvalues $E_{nk}^{\text{LDA}}$ using
\begin{equation}
\label{eq1}
 E_{nk}^{\text{QP}} = E_{nk}^{\text{LDA}} + Z(E_{nk}^{\text{LDA}}) \langle \psi_{nk} | \Sigma(E_{nk}^{\text{LDA}}) - V_{xc} | \psi_{nk} \rangle,
\end{equation}
where $V_{\text{xc}}$ is the LDA exchange-correlation potential, $\Sigma=iG_0W_0$ is the electronic self-energy evaluated at $E_{nk}^{\text{LDA}}$, $\psi_{nk}$ are one-particle Kohn-Sham states calculated with the LDA, and $Z(E_{nk}^{\text{LDA}})=\left[1-\frac{\partial \Re(\Sigma)}{\partial \omega} \big\rvert_{\omega=E_{nk}^{\text{LDA}}} \right]^{-1}$ is the QP renormalization factor (see Supporting Information for methodological and computational details). Consistent with prior studies, we find that the highest occupied states at the high symmetry point X of the Brillouin zone are primarily derived from Ag-$d_{z^2}$, Br-$p$, and Bi-$s$ orbitals, while the lowest unoccupied states at the same point are predominantly of Bi-$p$ and Br-$p$ character.

The calculated indirect band gap between the valence band maximum (VBM) at X and the conduction band minimum (CBM) at L is 1.66\,eV for Cs$_2$AgBiBr$_6$. This is slightly less than the range of experimentally reported band gaps of 1.8\,eV - 2.2\,eV\cite{McClure2016,Volonakis2016,Slavney2016a}, but consistent with previous GW calculations~\cite{Filip2016, Leppert2019l}. The smallest direct band gap is computed to be 2.41\,eV at X (Table~\ref{table1}). 

In addition, we calculate a hole effective mass of 0.31\,m$_0$ at X for Cs$_2$AgBiBr$_6$, similar to the values obtained from DFT and reported in the literature \cite{Volonakis2016}. In Table~\ref{table1} we show that the effective mass tensor is highly anisotropic, with the effective mass along the direction from X to $\Gamma$ at least four times smaller than those along the other two directions. This has been previously correlated with the rocksalt packing of Ag and Bi ions \cite{Slavney2019}.

\begin{table*}[htb]
    \centering
    \begin{tabular}{ccccccc} 
    \hline
    \multicolumn{1}{c}{} & \multicolumn{2}{c}{\textbf{ {QP} band gap (eV)}} & \multicolumn{4}{c}{\textbf{Hole effective mass (m$_0$)}} \\ \hline
 & X$^{\text{VBM}}$ $\to$ L$^{\text{CBM}}$ &  X$^{\text{VBM}}$ $\to$ X$^{\text{CBM}}$ & m$_{h_{_1}}$ & m$_{h_{_2}}$ & m$_{h_{_3}}$ &  m$_h^*$ \\ \hline
    DFT-LDA & 0.90 & 1.66 & 0.79 & 0.73 & 0.17 & 0.36 \\ 
    $G_0W_0$@LDA & 1.67 & 2.41 & 0.72 & 0.67 & 0.15 & 0.31 \\ \hline
\end{tabular}
    \caption{LDA and $G_0W_0$ lowest indirect and direct QP band gap and hole effective masses of cubic Cs$_2$AgBiBr$_6$ at X (in units of the electron rest mass m$_0$). The indices correspond to principal axes of the effective mass tensor. $m_{h{_3}}$ corresponds to the hole effective mass along the direction from X to $\Gamma$. $m_h^*$ is computed as the harmonic mean of the masses along the three principal components.}
    \label{table1}
\end{table*}

\begin{figure}[htb]
  	\centering
  	\includegraphics[width=0.9\columnwidth]{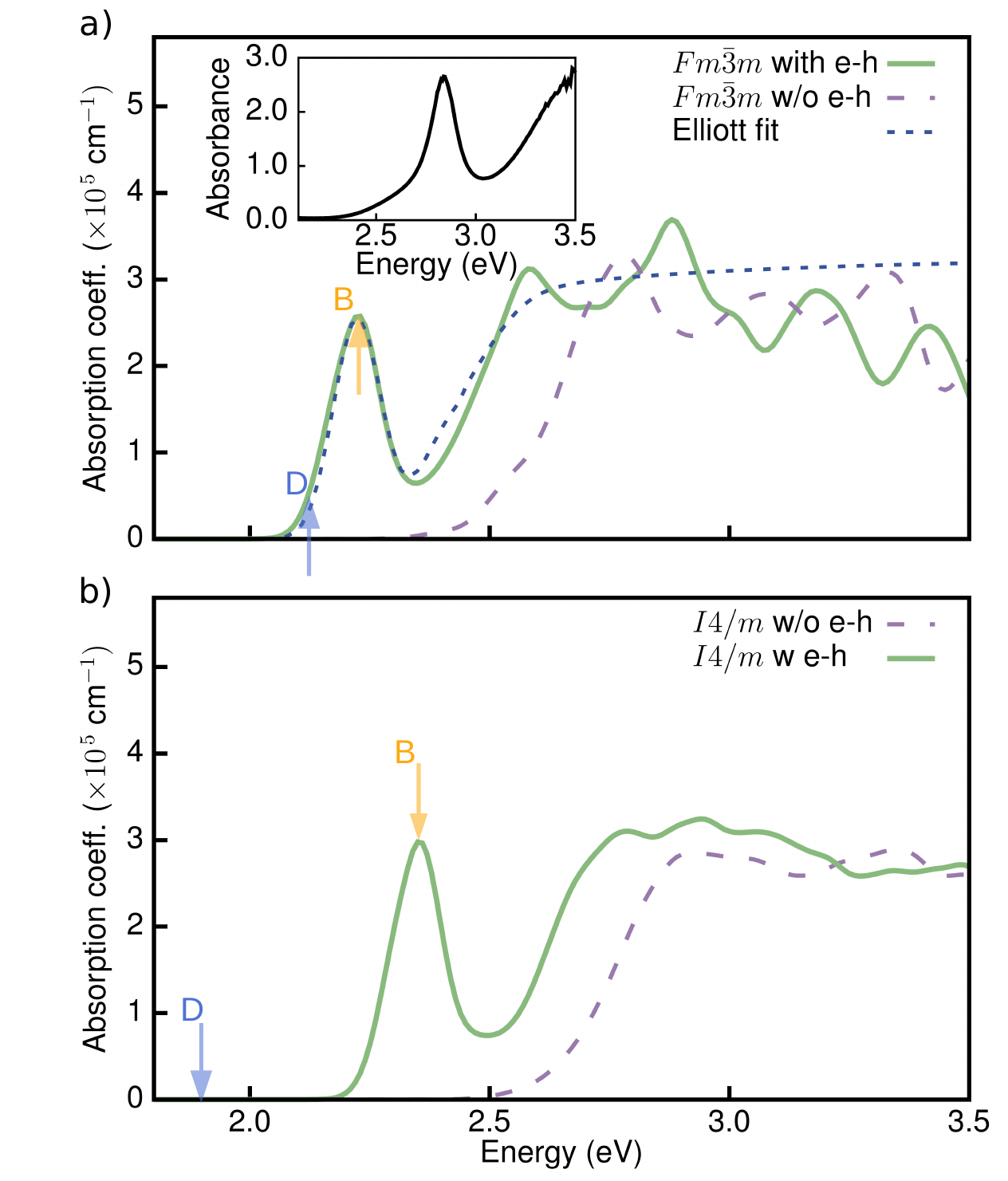}
  	\caption{a) Optical absorption spectrum of the cubic $Fm\bar{3}m$ phase of Cs$_2$AgBiBr$_6$, calculated using the random phase approximation (RPA) (purple dashed line), the GW+BSE approach (green solid line) and using the Elliott model (dark blue dashed line), experimental optical absorption spectrum with data from Ref.\citenum{Longo2020} (inset). b) Optical absorption spectrum of tetragonal $I4/m$ phase of Cs$_2$AgBiBr$_6$, calculated using the RPA (purple dashed line), the GW+BSE approach (green solid line). The blue (orange) arrow marks the first dark (bright) transition D (B). RPA optical spectra do not include local field effects.}
   	\label{fig2}
\end{figure}

Figure~\ref{fig2}a shows computed linear absorption spectra of cubic $Fm\bar{3}m$ Cs$_2$AgBiBr$_6$, obtained within the random phase approximation (RPA; without local field effects) and the $GW$+BSE approach~\cite{Rohlfing1998, Rohlfing2000}, i.e. without and with electron-hole interactions, respectively. Electron-hole interactions red-shift the absorption spectrum and give rise to a new sharp peak below the lowest direct band gap, but above the indirect band gap, indicative of a resonant exciton. A closer look at the fine structure of this peak, which is centered 570\,meV  above the indirect band gap, reveals a group of three degenerate bright states (marked as B in Figure~\ref{fig2}a). We also compute an optically inactive (dark) excitonic state (marked as D) 80\,meV below this peak. In the following, we will report the energy of the {\it first bright} exciton, unless stated otherwise, in order to aid comparison with experiment. Hereafter, we define the binding energy as the difference between the direct band gap and the computed excitation energies of these resonant bright excitons. 
Our computed exciton binding energy of 170~meV falls within the range of experimental values reported in the literature for cubic Cs$_2$AgBiBr$_6$ (between 70~\cite{Steele2018} and 268\,meV~\cite{Kentsch2018a}, (as determined from Elliott model fits), and it is a factor of two smaller than the value computed by Palummo et al.~\cite{Palummo2020} (340\,meV). Sources for the discrepancy between our calculations and Ref.~\citenum{Palummo2020} are the DFT starting point (for example, a DFT-PBE starting point increases the exciton binding energy somewhat, see SI), the density of the $\mathbf{k}$-point mesh (2.5-times denser in our case), and the use of partially self-consistent $GW$ in Ref.~\citenum{Palummo2020} (see SI for details). Despite the sensitivity of the band gap and exciton binding energy on the DFT starting point, the appearance of a well-defined peak as a consequence of the inclusion of electron-hole interactions is a robust result of our calculations.

We find a similar excitonic feature at the onset of the optical absorption spectrum in our calculations of the low-temperature tetragonal $I4/m$ phase \cite{Schade2019a}, but the entire spectrum is blue-shifted with respect to the cubic $Fm\bar{3}m$ phase by $\sim$150~meV, as shown in Figure~\ref{fig2}b. For the $I4/m$ phase, this is consistent with the slightly larger direct band gap of 2.54~eV (see Figure~S4), and in line with the general observation that octahedral tilting induces larger band gaps in halide perovskites~\cite{Filip2014}. Furthermore, we find a range of dark excitonic states up to 414\,meV below the first bright state, which can be associated with band folding in the $I4/m$ phase (marked as D and B, respectively, in Figure~\ref{fig2}b). These results hint that the experimentally-observed photoluminescence $\sim$1\,eV below the absorption onset may be related to phonon-assisted optical transitions \cite{Dey2020}, although further investigations would be required to confirm this.
 
To further validate our calculations, we compare our calculated optical spectra with the experimental optical absorption spectrum reported by Longo et al.~\cite{Longo2020}, shown in the inset of Figure~\ref{fig2}a. Our calculated optical absorption spectrum is red-shifted with respect to experiment by approximately 0.6 eV. This underestimation is consistent with prior studies of the QP band gaps in both halide double perovskites and lead-based perovskites and can be understood as originating with limitations associated with the DFT starting point~\cite{Filip2014, Leppert2019l}. Nonetheless, the experimental and theoretical lineshapes are very similar, and exhibit a well-defined peak before the onset of a broader continuum; our GW+BSE calculations allow us to assign this peak to a resonant excitonic feature.
 
It is instructive to analyze our computed spectrum using Elliott theory~\cite{Elliott1957}, a standard phenomenological theory of hydrogenic excitons in solids, typically used to extract exciton binding energies and band gaps from experimental optical absorption spectra. By fitting our BSE spectrum with the Elliott formula for the optical absorption coefficient as described in Ref.~\citenum{Davies2018}, from 0.2\,eV below the onset up to 1.2\,eV above the onset, we obtain an exciton binding energy of 231\,meV, the same  order of magnitude but $\sim$35\% higher than our BSE calculations. This comparison illustrates that the Elliott theory does not fully capture the nature of electron-hole interactions in this system, and that the excitons do not obey the hydrogenic Wannier-Mott model~\cite{Wannier-exc}. We compute the Wannier-Mott binding energy using the expression $\displaystyle{E_x=\frac{\mu}{\varepsilon_\infty^2}R_H}$, where $R_H$ is the Rydberg constant, and values of the reduced effective mass $\mu$ and dielectric constant $\varepsilon_\infty$ are obtained from our $GW$ calculations. Note, that the low dispersion of the CBM at X introduces a large uncertainty in the calculation of the electron effective mass (Table S2). We therefore approximate $\mu$ with the orientationally-averaged hole effective mass (see SI for details). With this approximation, the Wannier-Mott model underestimates the exciton binding energy for Cs$_2$AgBiBr$_6$ by $\sim$30\%. Inclusion of the electron effective mass would lower the Wannier-Mott exciton binding energy further, to less than half of our first principles result. The non-hydrogenic nature of the exciton also becomes apparent by inspecting the GW+BSE binding energy of the second excited state (Table~\ref{table3}), which deviates significantly from the hydrogenic Rydberg series expected based on the Wannier-Mott model. Even when effective mass anisotropy is considered, the Wannier-Mott model underestimates the binding energy of the second excited state by $\sim$150\%.
\begin{table*}[htb]
    \centering
  		\begin{tabular}{cccc}
  			\hline
\textbf{n} & \textbf{$G_0W_0$@LDA+BSE} & \textbf{Wannier-Mott}  & \textbf{Wannier-Mott with} $m^*(\lambda)$ \\ \hline
1 & 170                    & 120                & 148  \\
2 & 93                     & 30                 & 37    \\ \hline
  	    \end{tabular}%
  	      \caption{Exciton binding energies (in meV) of the two lowest energy bright states as calculated with $G_0W_0$@LDA+BSE, the Wannier-Mott model, and the Wannier-Mott model including effective mass anisotropy (see main text).}
  \label{table3}
\end{table*}

We assign the misalignment of the Wannier-Mott model and first principles results to the anisotropic QP band structure of these halide double perovskites -- in particular, the differences between the longitudinal and transverse effective masses are significant (Table~\ref{table1}) -- and to local field effects of the dielectric function. To probe the former, we use a hydrogenic model expression that explicitly accounts for effective mass anisotropy, as presented in Ref.~\citenum{Schindlmayr1997} (see SI and Table S3 for details). We find that including the effective mass anisotropy increases the Wannier-Mott binding energy by $\sim$23\% for Cs$_2$AgBiBr$_6$, bringing the model closer to the \textit{ab initio} result (Table~\ref{table3}). To probe the latter, we artificially modify the first principles dielectric function such that  $\epsilon(\mathbf{r}, \mathbf{r'};\omega) =  \varepsilon_\infty$, and obtain an exciton binding energy underestimated with respect to the first principles result by $\sim$42\%. The importance of local field effects can also be seen in the imaginary part of the dielectric function, where they lead to a significant suppression of the spectrum when electron-hole interactions are \textit{not} considered (see Figure S5), consistent with prior studies~\cite{Louie1975,Puschnig2002}.

\begin{figure}[htb]   
  	\centering
  	\includegraphics[width=0.9\columnwidth]{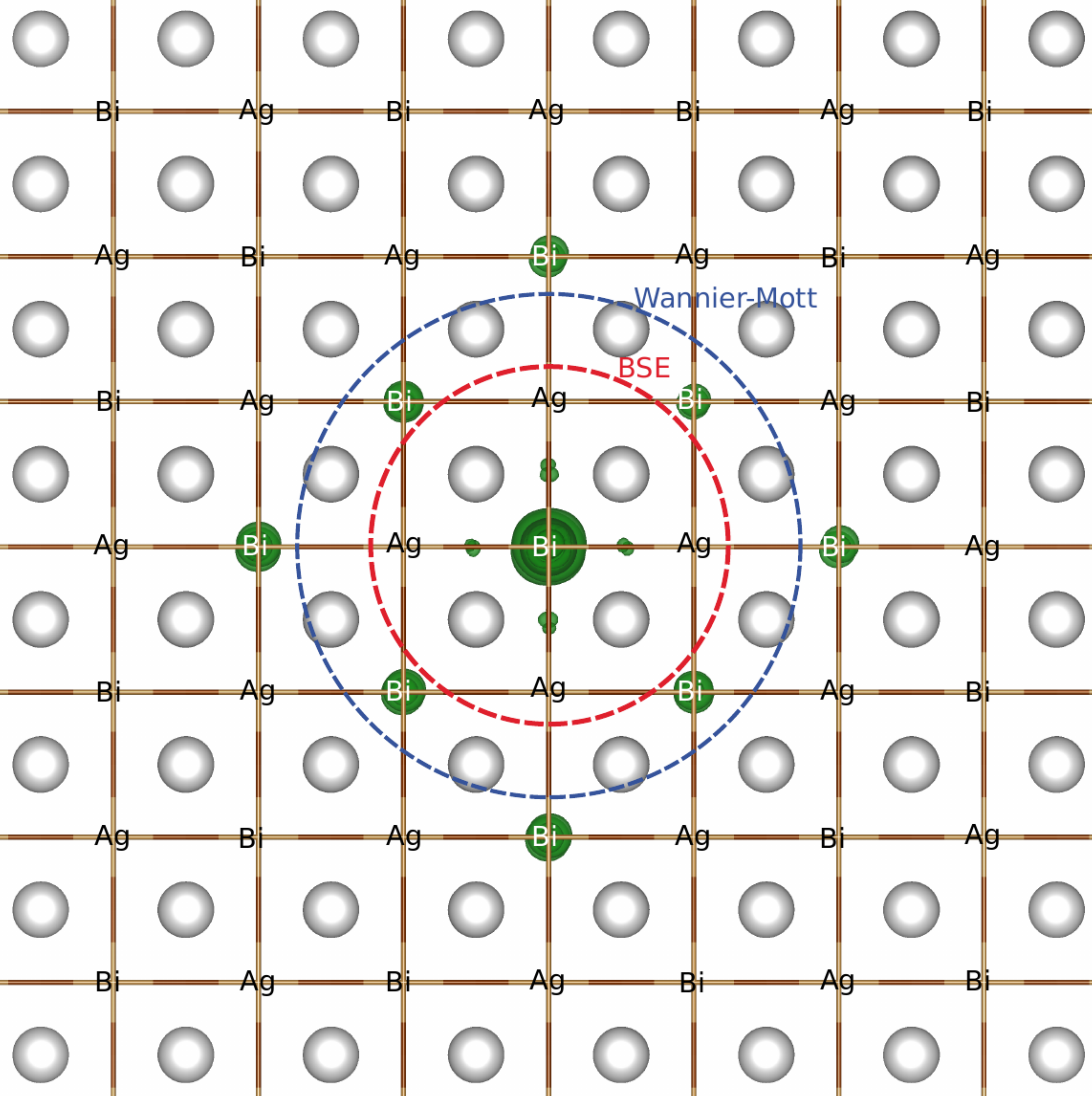}
  	\caption{3D representation of the probability density of the exciton wave function in real space (depicted as green isosurfaces), showing 95\% of the maximum isovalue. The position of the hole is fixed on a Bi ion, and the circles represent the average electron-hole separation as computed from first principles (red) and the Wannier-Mott model (blue). The silver spheres are Cs.}
   	\label{fig3}
\end{figure}

Excitons with a large binding energy are expected to be highly-localized within the crystal lattice. In Figure~\ref{fig3} we show the probability distribution of the excitonic wave function $\Psi_S(\mathbf{r}_e,\mathbf{r}_h)=\sum_{vc\mathbf{k}}A^S_{vc\mathbf{k}}\psi_{c\mathbf{k}}(\mathbf{r}_e)\psi^*_{v\mathbf{k}}(\mathbf{r}_h)$, where $\psi_{v(c)\mathbf{k}}(\mathbf{r}_{h(e)})$ are single-particle DFT Kohn-Sham wave functions for the electrons and holes, and $A^S_{vc\mathbf{k}}$ are coefficients corresponding to the excitonic state $S$, calculated directly from the BSE (see SI for details). To visualize $\Psi_S(\mathbf{r}_e,\mathbf{r}_h)$, we fix the position of the hole at a Bi site (see SI for details), and plot the distribution of $|\Psi_S(\mathbf{r}_e,\mathbf{r}_{\text{Bi}})|^2$ in real space. Figure~\ref{fig3} shows that the probability density of the exciton wave function departs significantly from a spherical hydrogenic probability distribution. Analysis of the coefficients $A^S_{vc\mathbf{k}}$ reveals that more than 90\% of the excitonic wavefunction is comprised of VBM $\rightarrow$ CBM transitions at X with anisotropic (Ag-$d_{z^2}$/Bi-$s$/Br-$p$ $\rightarrow$ Bi-$p$/Br-$p$, see Figure~\ref{fig3}) orbital character, highlighting the heterogeneous and anisotropic nature of the exciton. Following the approach of Ref.~\citenum{Sharifzadeh2013}, we quantify the exciton's spatial extent by computing the average electron-hole separation denoted by $\displaystyle{\sigma_\mathbf{r} = \sqrt{\langle|\mathbf{r}|^2\rangle - \langle |\mathbf{r}|\rangle^2}}$ where 
\begin{equation}
    \langle|\mathbf{r}|^n\rangle = \int_\Omega d^3\mathbf{r} \abs{\mathbf{r}}^n F_S(\mathbf{r})   
\end{equation}
is the $n$-th moment of the electron-hole correlation function $F_S(\mathbf{r})$ and $\Omega$ the volume of the supercell (see SI for details). We find that the excitonic wave function is highly localized, with $\sigma_\mathbf{r} = 6.3$\,\AA. In Figure~\ref{fig3}, we compare this result with the Wannier-Mott model by computing the average electron-hole separation of a hydrogenic wavefunction ($\displaystyle{\sigma_\mathbf{r}^{\text{WM}} = \frac{\sqrt{3}}{2}a_H}$, where $a_H$ is the Bohr radius). We find that $\sigma_\mathbf{r}^{\text{WM}}=8.76$\,\AA, $\sim$37\,\% larger than the result from our more rigorous BSE calculation.

\begin{figure}[htb]
  	\centering
  	\includegraphics[width=0.9\columnwidth]{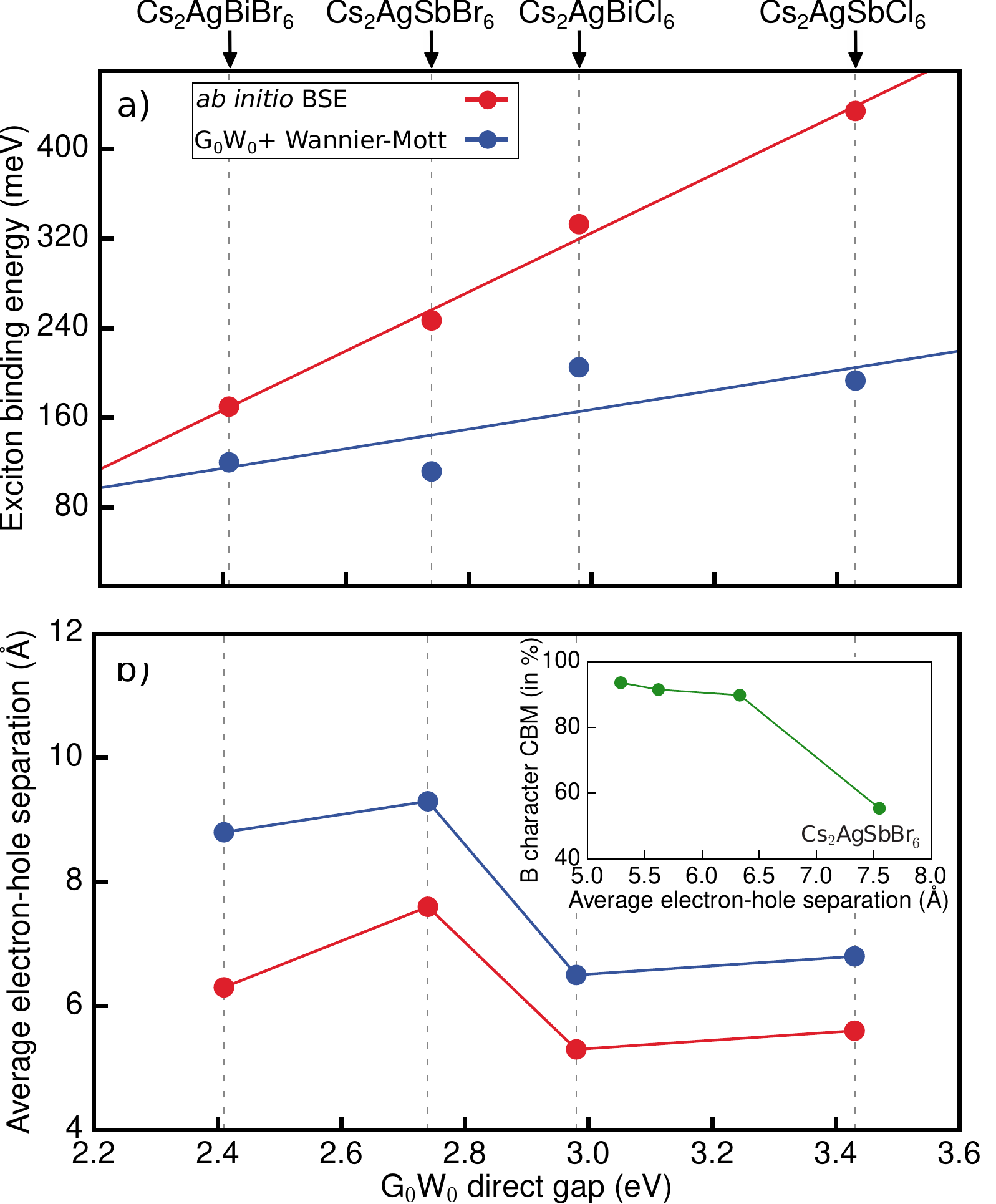}
  	\caption{a) Variation of the exciton binding energy as computed within the BSE approach (red) and the Wannier-Mott model (blue) with respect to the G$_{0}$W$_{0}$ lowest direct band gap. b) average electron-hole separation as computed with BSE (red) and as predicted by the Wannier-Mott model (blue) with respect to the G$_{0}$W$_{0}$ lowest direct band gap. Inset: Percentage B orbital character of the CBM with respect to average electron-hole separation as computed with the BSE approach.}
  	\label{fig4}
\end{figure}

To understand how the chemical composition impacts the excitonic properties of halide double perovskites, we extend our calculations and analysis to other members of the Ag-pnictogen halide perovskite family, namely cubic Cs$_2$AgBiCl$_6$, Cs$_2$AgSbBr$_6$, and Cs$_2$AgSbCl$_6$. {The results of our first principles calculations are summarised in  Table \ref{table2}}. As shown in Figure~\ref{fig4}a, the Wannier-Mott model severely underestimates the exciton binding energies relative to our GW+BSE calculations for all compounds, which range between 170 and 434\,meV and scale linearly with the lowest direct band gap. Furthermore, the Wannier-Mott model also predicts a linear dependence of the binding energy on the lowest direct band gap, albeit with a different slope than our GW+BSE calculations, consistent with the variation of the effective mass anisotropy among these materials (Table S3). Furthermore, as shown in Figure~S6, the Elliott model fails to describe optical absorption lineshapes for this entire family of double perovskites. 

  \begin{table*}[htb]
    \centering
  		\begin{tabular}{@{}ccccccc@{}}
  			\hline
  			\multirow{2}{*}{\textbf{ }} & \multicolumn{2}{c}{\textbf{Lowest direct gap (eV)}} & \multirow{2}{*}{\textbf{\begin{tabular}[c]{@{}c@{}}Exciton binding \\ energy (meV)\end{tabular}}}&$\Huge \varepsilon_\infty$ & \multirow{2}{*}{\textbf{\begin{tabular}[c]{@{}c@{}}Average \\ e-h separation (\AA)\end{tabular}}}\\
  			& \textbf{DFT-LDA}                 & \textbf{$G_0W_0$@LDA}                &                                                                                                   \\ \hline
  			Cs$_2$AgBiBr$_6$                 & 1.67                         & 2.41                        & 170  & 5.92 & 6.3                                                                                              \\
  			Cs$_2$AgBiCl$_6$                 & 1.89                         & 2.98                        & 333  & 4.68 & 5.3                                                                                               \\
  			Cs$_2$AgSbBr$_6$                 & 1.79                         & 2.74                        & 247  & 5.96 & 7.6                                                                                               \\
  			Cs$_2$AgSbCl$_6$                 & 2.28                         & 3.43                        & 434  & 4.77 & 5.6                                                                          
  		                      \\ \hline
  	    \end{tabular}%
  \caption{DFT-LDA and $G_0W_0$@LDA lowest direct transition (in eV), exciton binding energy (in meV),  static dielectric constant as computed within the random phase approximation, and average electron-hole separation. }
  \label{table2}
\end{table*}

In line with the large binding energies, exciton localization is a common feature of all four compounds in this series (see Figure~\ref{fig4}b). However, the degree of exciton localization we compute does not follow the trends expected from the Wannier-Mott model. Specifically, our calculations show the strongest exciton localization for Cl-based double perovskites, with an average electron-hole separation just slightly larger than one unit cell (see Figure S7). Surprisingly, Cs$_2$AgSbBr$_6$ exhibits a more delocalized exciton than Cs$_2$AgBiBr$_6$ (7.6\,\AA~vs.~6.3\,\AA , respectively) even though its exciton binding energy of 247\,meV is significantly higher than that of Cs$_2$AgBiBr$_6$. We hypothesize that the exciton localization is correlated with the spatial extent of the electronic states from which the CBM is derived. Indeed, the inset of Figure~\ref{fig4}b shows that the average electron-hole separation scales with the fractional contribution of the B-$p$ character of the CBM, where B=Bi or Sb. The greater the B-$p$ orbital character of the CBM is, the more strongly the exciton is localized. Based on this trend, we can understand the comparably greater de-localization of the excitonic wave function of Cs$_2$AgSbBr$_6$ as a consequence of the reduced Sb-$p$ contribution at the CBM (see Figure~S8). In fact, Cs$_2$AgSbBr$_6$ is the only case for which the lowest energy direct band gap is located at L, and the exciton is primarily derived from interband transitions at L. The CBM at L has pronounced Ag-$s$ orbital character, which leads to a more isotropic effective mass and delocalized exciton in this case. As before, we also calculate the average electron-hole separation based on the Wannier-Mott model (Figure~\ref{fig4}b) and find that this value systematically overestimates our first principles results.

The results of our calculations of Ag-pnictogen halide double perovskites have uncovered new intuition for the physics of excitons in this family of materials. Firstly, resonant excitons and optical spectra depart significantly from the Elliott model, which assumes a direct band gap semiconductor with parabolic band edges, isotropic effective masses, and weakly bound Wannier-Mott-like excitons~\cite{Elliott1957}. Secondly, excitons in these 3D double perovskite crystals exhibit large binding energies, of similar magnitude to those observed in quantum confined systems\cite{Blancon2018,Rieger2019}. This is especially notable given that the closely related 3D lead-halide perovskites have similar band gaps but exhibit weakly bound excitons which easily dissociate at room temperature~\cite{Bokdam2016}. Replacing Pb with Ag and Bi or Sb at the B site results in the electronic charge density becoming less uniform and isotropic, and electronic states at the band edges becoming more localized (or confined) within individual octahedra. In Bi- and Sb-based halide double perovskites, this chemical ``confinement'' appears due to the localization of electrons and holes in chemically distinct octahedra. 

In conclusion, we performed a systematic first principles study of the electronic and optical properties of the Cs$_2$AgBX$_6$ perovskite series (B = Bi, Sb and X = Cl, Br), comparing where possible with optical absorption experiments. We have shown that this family of halide double perovskites features strongly localized resonant excitons, with energies of up to $\sim$450\,meV below the direct band gap. In particular, Cs$_2$AgBiBr$_6$, a non-toxic, stable candidate material for replacing Pb-based halide perovskites, has a calculated exciton binding energy of $\sim$170\,meV, significantly higher than that of the closely related lead-halide perovskites. We demonstrated that the excitonic properties of these double perovskites are not well described by the Wannier-Mott hydrogenic model and their optical spectra do not obey the Elliott model, both methods routinely used for extracting exciton binding energies from experimental optical absorption spectra. Our results demonstrate how newly designed lead-free halide double perovskites have the potential to challenge conventional intuition and  understanding of light-matter interactions in chemically heterogeneous semiconductors.

\begin{acknowledgements}
The authors would like to acknowledge Giulia Longo (Northumbria U.) and Laura Herz (Oxford U.) for useful discussions and for sharing the data published in Ref.~\citenum{Longo2020}.
This work was supported by Theory FWP at the Lawrence Berkeley National Laboratory, which is funded by the U. S. Department of Energy, Office of Science, Basic Energy Sciences, Materials Sciences and Engineering Division, under Contract No. DE-C02-05CH11231. The authors would like to acknowledge computational resources at the Molecular Foundry, also supported by the Office of Science, Office of Basic Energy Sciences, of the US DOE under Contract DE-AC02-5CH11231 and resources of the National Energy Research Scientific Computing Center (NERSC). L.L. and R.B. were supported by the Bavarian State Ministry of Science and the Arts through the Collaborative Research Network Solar Technologies go Hybrid (SolTech), the Elite Network Bavaria, and the German Research Foundation (DFG) through SFB840 B7, and through computational resources provided by the Bavarian Polymer Institute (BPI). R.B. acknowledges support by the DFG program GRK1640. M.R.F. acknowledges support of the John Fell Oxford University Press (OUP) Research Fund.
\end{acknowledgements}

\end{document}